\documentclass[twoside]{dis09}
\usepackage[latin1]{inputenc}
\usepackage[dvips]{graphicx,epsfig,color}
\usepackage{wrapfig,rotating}
\usepackage{amssymb,amsmath,array}

\begin{document}
\title{Recent Results from the H1 Collaboration}

\author{Stefan Schmitt$^1$
%
\thanks{On behalf of the H1 Collaboration}
%
\vspace{.3cm}\\
%
DESY Hamburg \\
Notkestra\ss{}e 85, 22607 Hamburg - Germany
%
}

\maketitle

\begin{abstract}
An overview of recent physics results from the H1 collaboration is
given. The covered areas are: 
rare processes and searches for new physics, structure functions
and inclusive measurements, heavy flavour production, QCD and hadronic
final states, diffractive scattering.
\end{abstract}

\section{Introduction}

During the years 1994--2007 the HERA $ep$ collider in Hamburg collided
electrons or positrons with an energy of $27.5\,\text{GeV}$  and protons with
an energy of $920\,\text{GeV}$, at a centre-of-mass energy of
$320\,\text{GeV}$\footnote{A small fraction of the data was taken with a
  proton energy of $820\,\text{GeV}$.}. The last months of data taking were
dedicated to special low energy runs, with proton energies of $460$ and
$575\,\text{GeV}$, corresponding to centre-of-mass energies of $220$ and
$250\,\text{GeV}$, respectively. During the whole data taking period, the H1
detector recorded almost $200\,\text{pb}^{-1}$ in $e^{-}p$ collisions and
almost $300\,\text{pb}^{-1}$ in $e^{+}p$ collisions at high
centre-of-mass energy. These data are now fully
exploited to achieve milestones of the HERA physics program. A few
recent results are presented here \cite{url}.

\unitlength\columnwidth
\section{Rare processes and searches for new physics}

\begin{figure}[t]
\begin{center}
\includegraphics[width=0.45\unitlength]{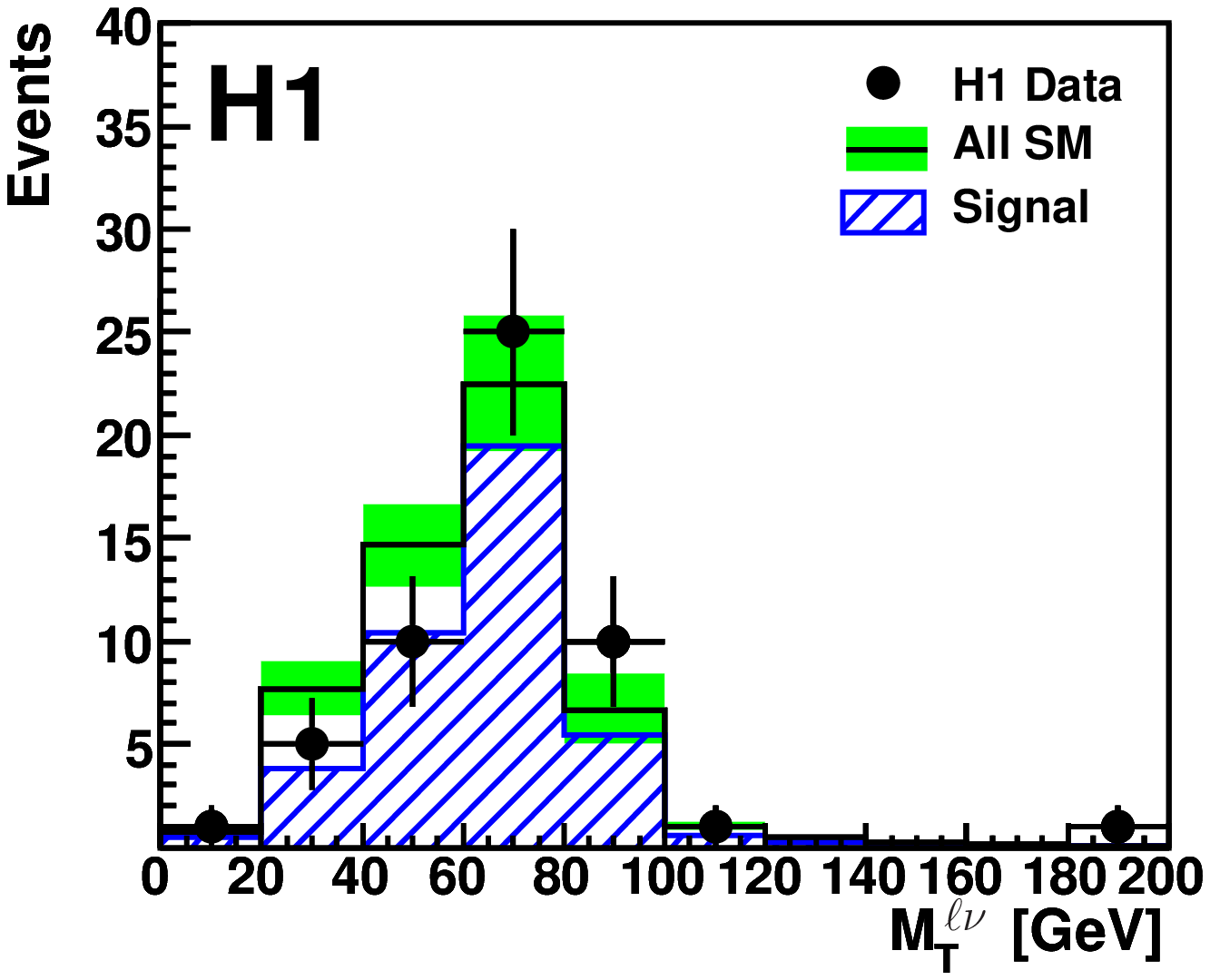}
\includegraphics[width=0.49\unitlength]{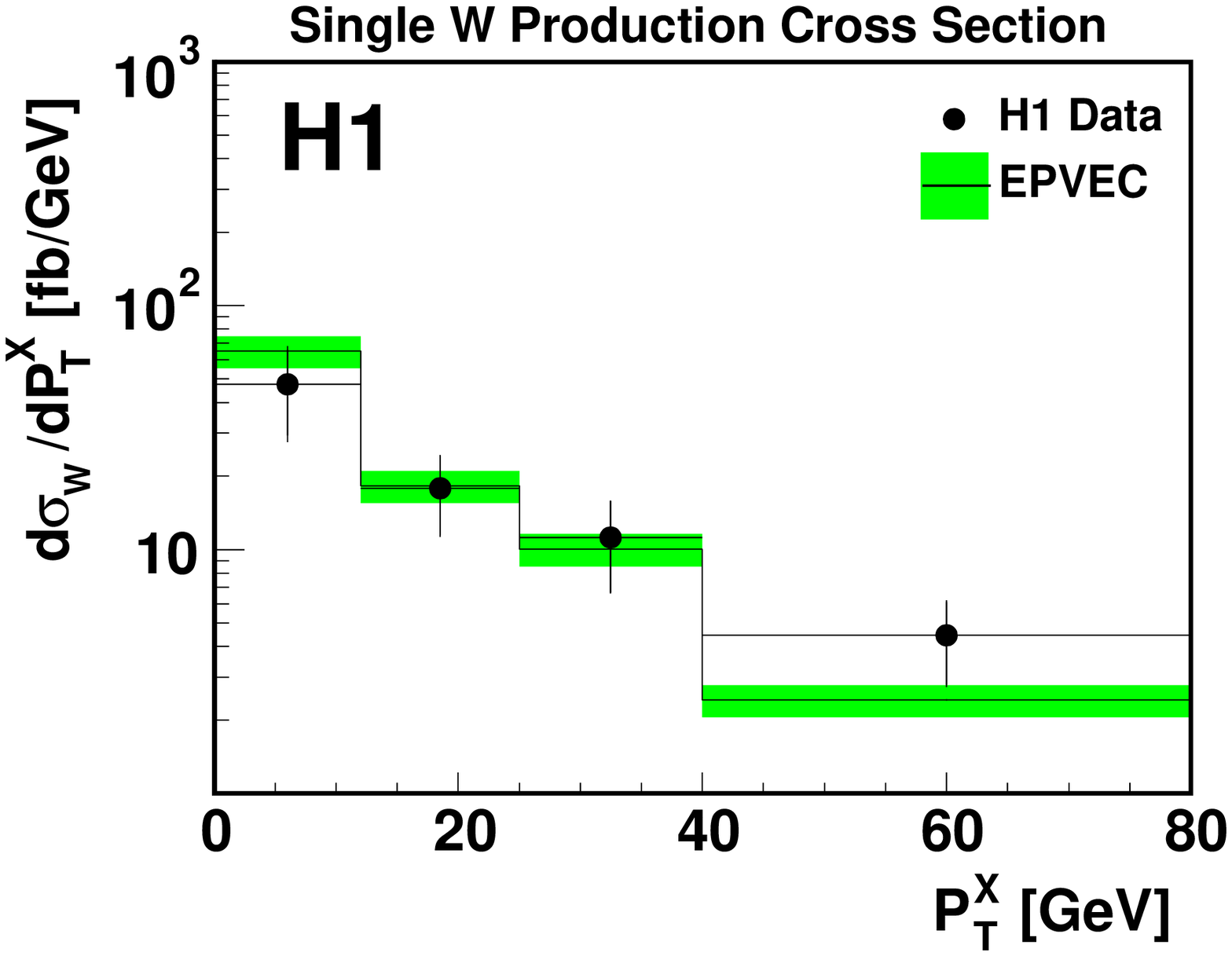}
\caption{Single $W$ production at HERA}\label{fig:isolep}
\end{center}
\end{figure}

At HERA, $W$ bosons may be singly produced in the reaction $e^{\pm}p\to
e^{\pm}WX$. Leptonic decays of the $W$ bosons lead to the spectacular
signatures of a high momentum isolated lepton and missing transverse momentum
due to the decay neutrino. Such events are measured by H1
\cite{Aaron:2009wp} as a function of the transverse momentum of the
hadronic system $P_T^X$.
At large $P_T^X>25\,\text{GeV}$ an excess of
$17$ observed over $8.0\pm1.3$ expected events is seen in
$e^{+}p$ scattering.  In $e^{-}p$ scattering only $1$ event is seen,
with $5.6\pm0.9$ expected from the Standard Model (SM).
The transverse mass distribution of the isolated lepton and the neutrino is
exhibits a shape with a falling edge above $80\,\text{GeV}$, as expected from
$W$ decays. The production cross-section of single $W$ at HERA is measured as
a function of $P_T^X$ and is shown in Figure \ref{fig:isolep}, together with
the observed transverse mass distribution. A total cross-section of
$1.14\pm0.25\pm0.14\,\text{pb}^{-1}$ is found.

\begin{figure}[t]
\begin{center}
\includegraphics[width=0.9\unitlength]{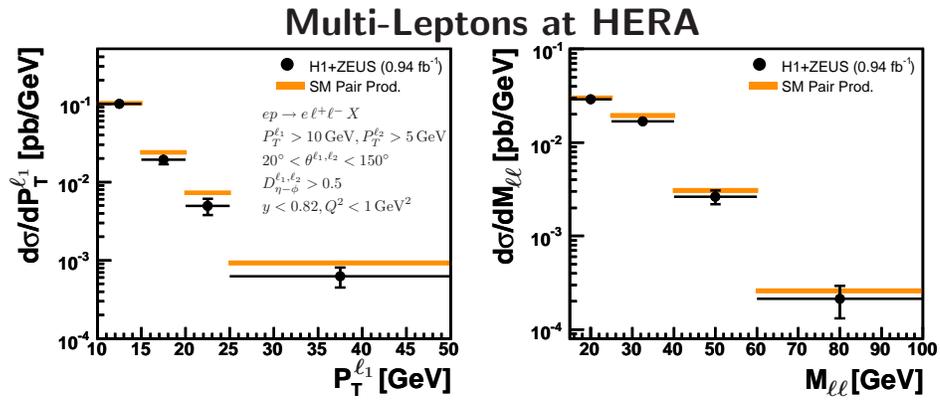}
\caption{Multi-lepton production at HERA}\label{fig:multilep}
\end{center}
\end{figure}

Events with several high $P_T$ leptons ($e$ or $\mu$) are also
investigated. Within the SM such events are
predominantly produced from $\gamma\gamma$ collisions. The H1 and ZEUS
collaboration both measure individually these topologies
\cite{Aaron:2008jh,Collaboration:2009cv}. In order to enhance the 
sensitivity to new phenomena, a combined H1 and ZEUS analysis is
performed \cite{Collaboration:2009sm} in a common phase-space. Multi-lepton events are observed at high scalar sum of transverse momentum $\sum p_T>100\,\text{GeV}$. In $e^{+}p$ collisions, $7$ events are observed for $1.94\pm0.17$ expected, whereas in $e^{-}p$ collisions no event is seen, for $1.19\pm 0.12$ expected. The cross-section for multi-lepton production is measured as a function of the scalar sum of the lepton
$P_T$ and as a function of the invariant mass, as depicted in Figure
\ref{fig:multilep}.

\begin{figure}[b]
\begin{center}
\begin{picture}(1,0.27)
\put(0.0,0){\includegraphics[width=0.33\unitlength]{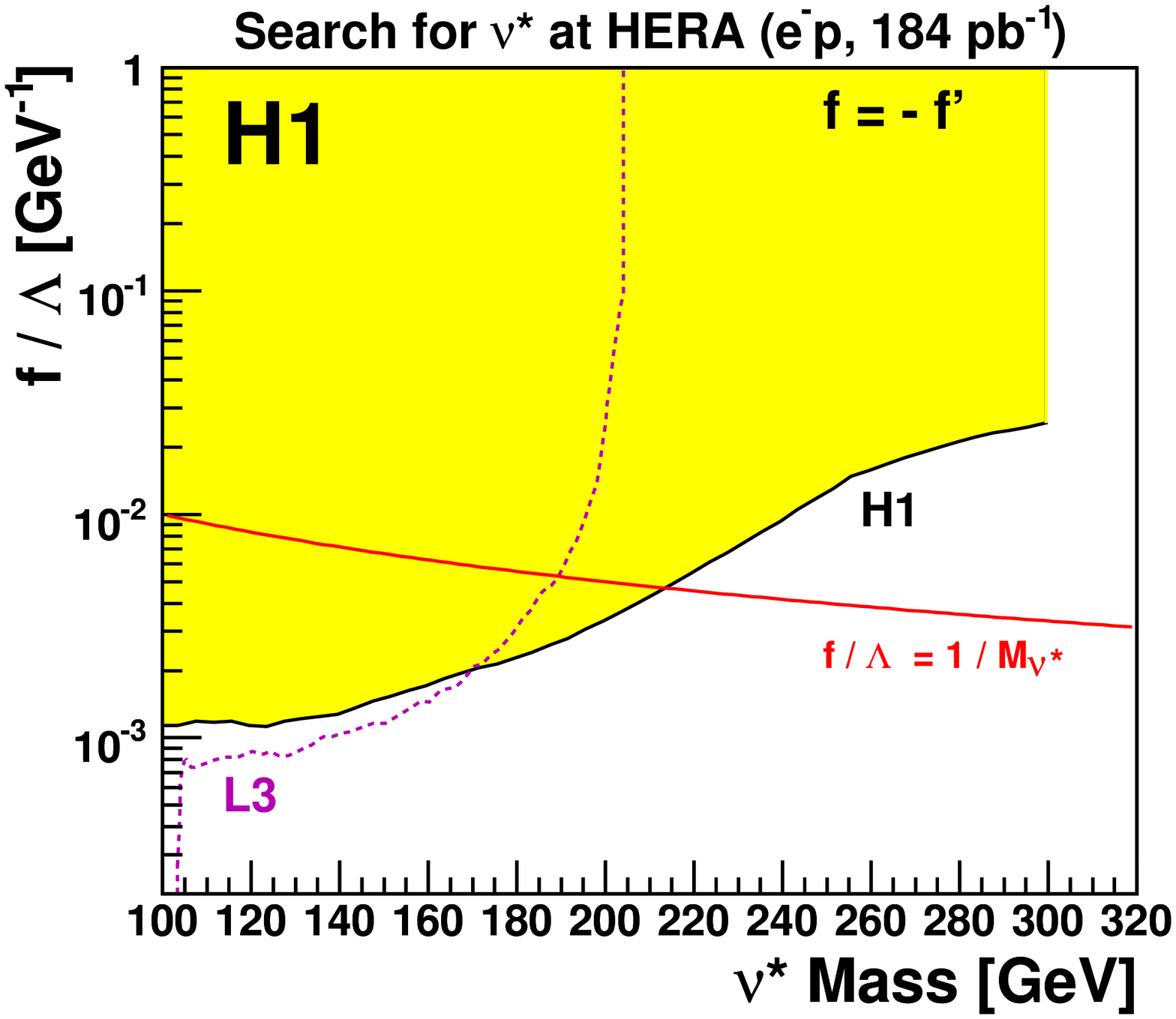}}
\put(0.333,0){\includegraphics[width=0.33\unitlength]{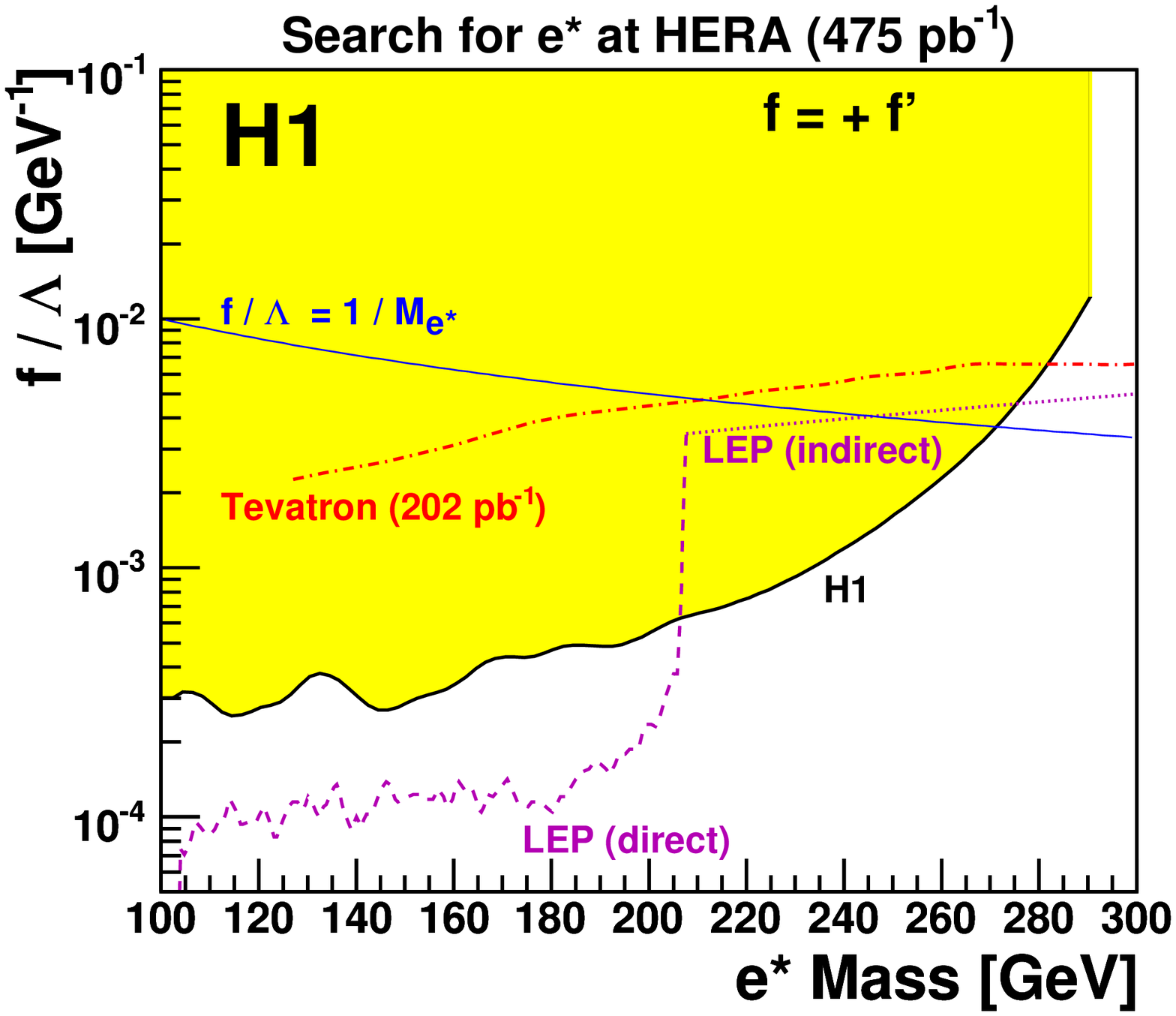}}
\put(0.667,0){\includegraphics[width=0.33\unitlength]{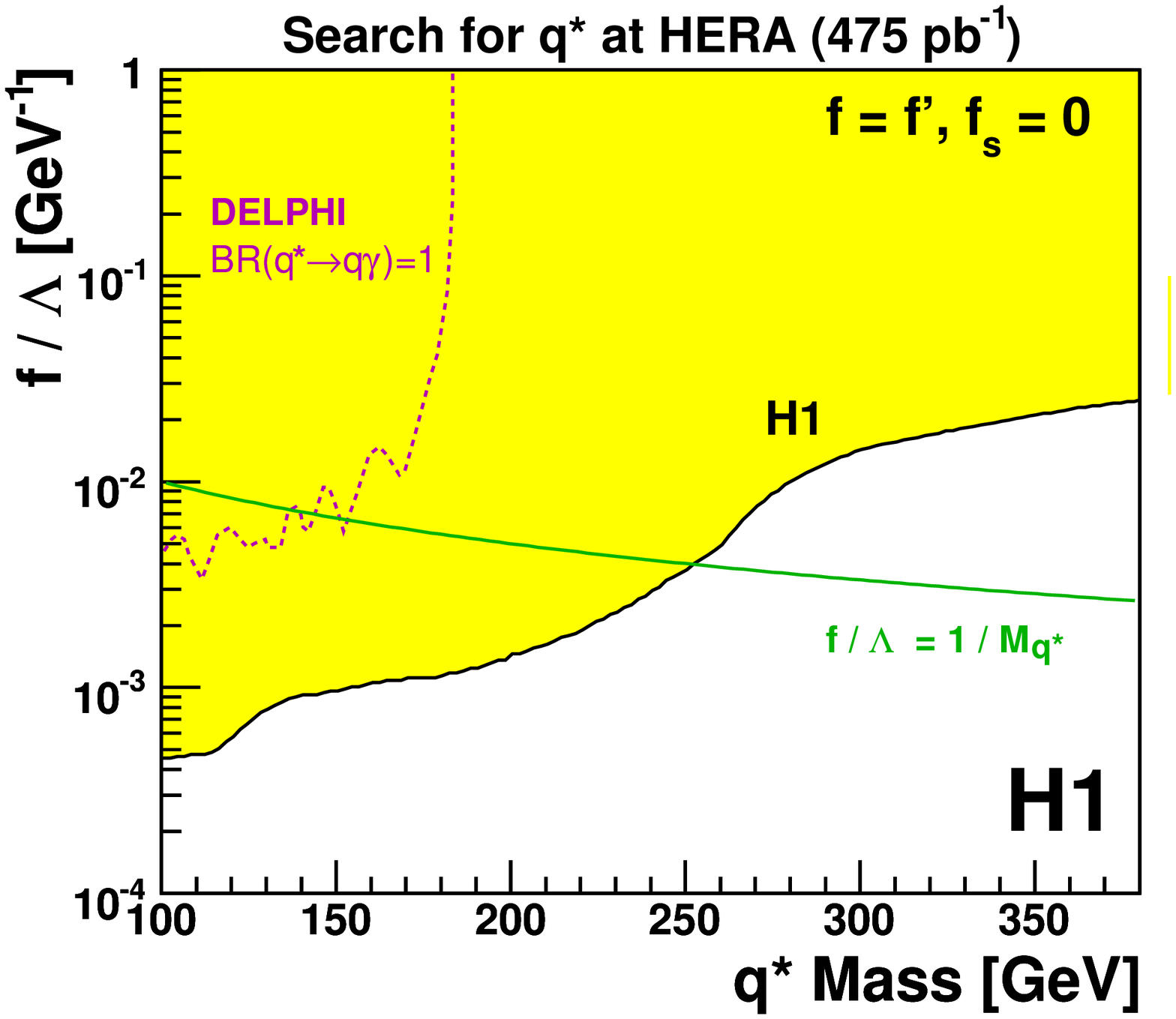}}
\end{picture}
\caption{Exclusion limits on excited fermion production at HERA}\label{fig:excited}
\end{center}
\end{figure}
The H1 data also have been searched for excited
fermions. Mass-dependent limits on the coupling $f/\Lambda$ for the production
of excited neutrinos \cite{:2008xe}, excited electrons \cite{Aaron:2008cy} and
excited quarks \cite{Collaboration:2009iz} are shown in Figure
\ref{fig:excited}. The excited quark limits are derived with the
additional assumption $f_s=0$, complementary to TeVatron searches,
where $f_s=f$ is assumed. Using the conventional assumption
$\Lambda/f=m_{f^\star}$, masses $m_{\nu^\star}<213\,\text{GeV}$,
$m_{e^\star}<272\,\text{GeV}$ and $m_{q^\star}<252\,\text{GeV}$ are excluded at $95\%$ CL.

\section{Structure functions and inclusive measurements}

\begin{figure}[b]
\begin{center}
\includegraphics[width=0.72\columnwidth]{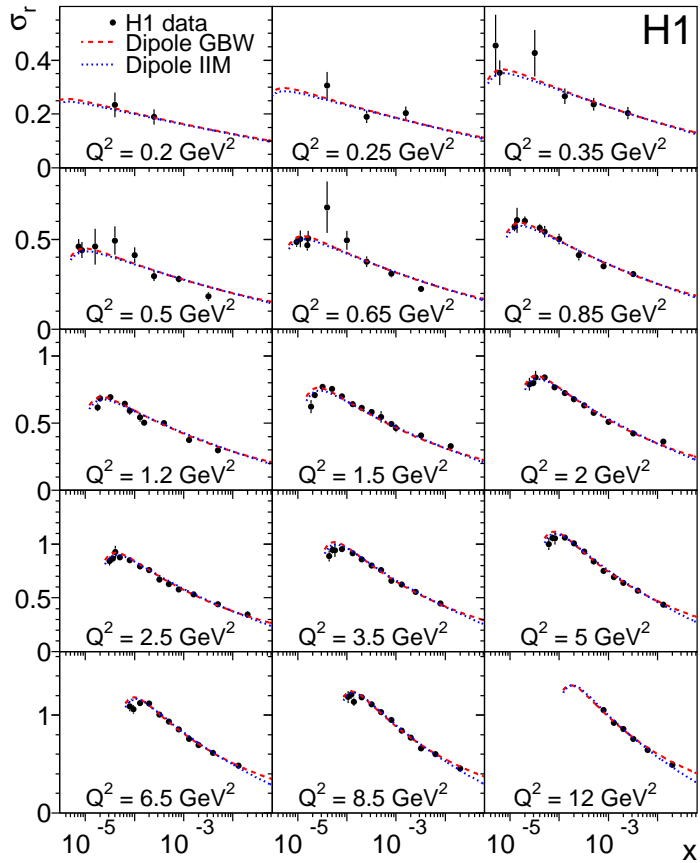}
\caption{The reduced cross-section $\sigma_r$ at low $Q^2$}\label{fig:lowq2}
\end{center}
\end{figure}
\begin{figure}[t]
\begin{center}
\includegraphics[height=0.4\columnwidth]{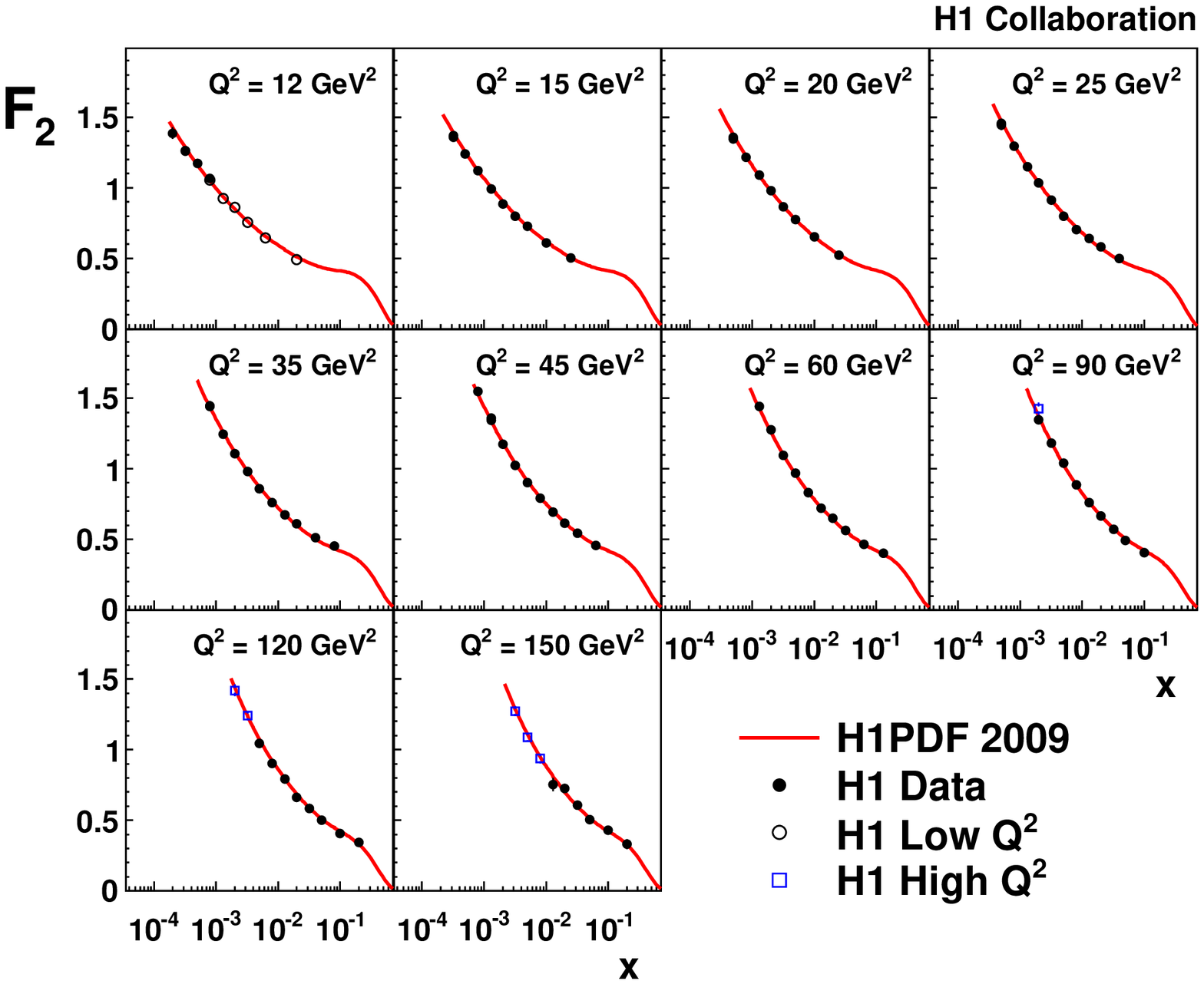}
\includegraphics[height=0.4\columnwidth]{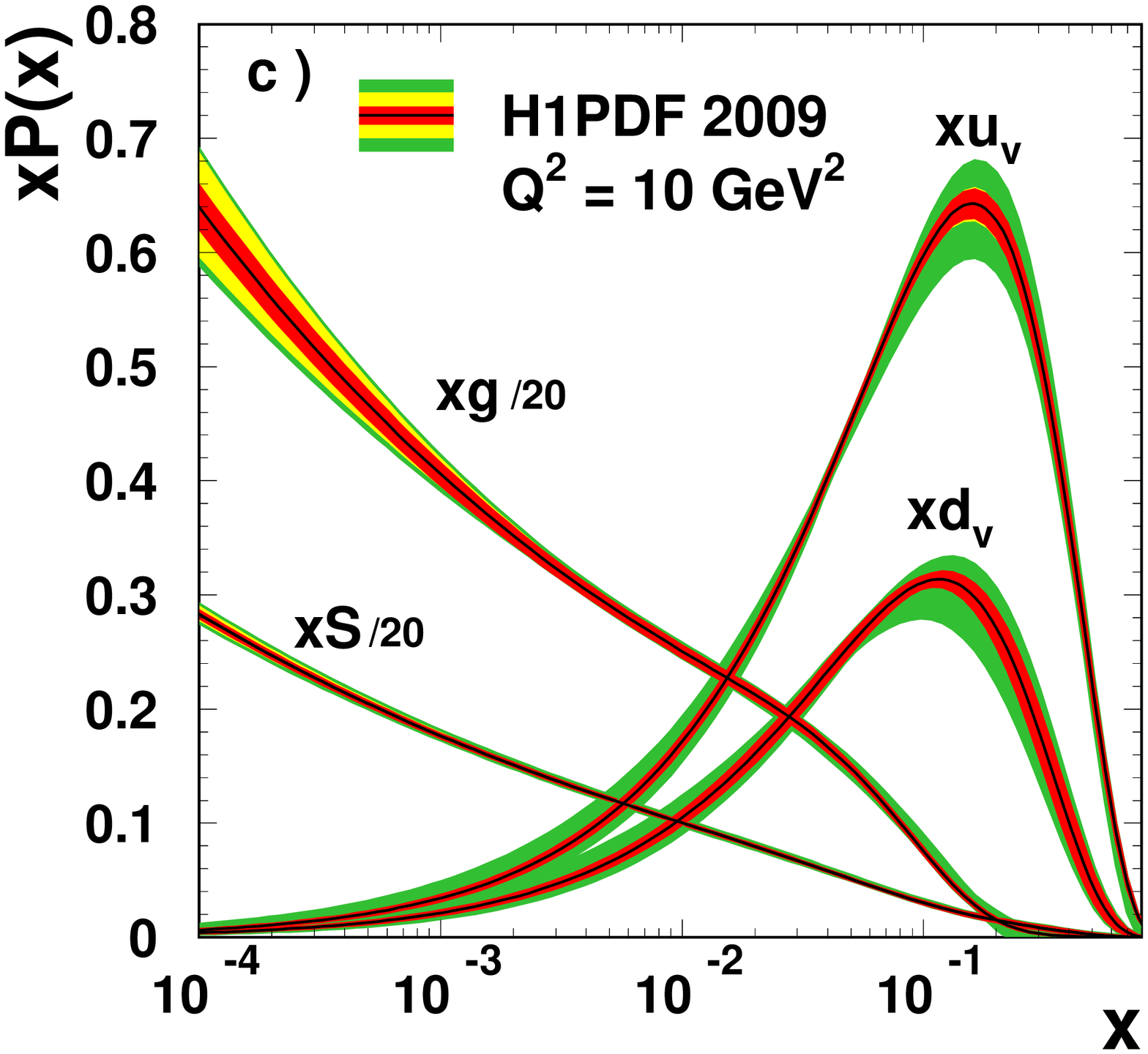}
\caption{The structure function $F_2$ (left) and the parton
  distribution functions obtained from the H1PDF2009 fit (right)}\label{fig:inclusive}
\end{center}
\end{figure}

The measurement of structure functions is one of the key elements of
HERA physics. Recent analysis from H1 reach unprecedented accuracy
in this area. The data at low momentum transfer
$Q^2\lesssim12\,\text{GeV}$, have a precision of $2-3$ percent 
\cite{Collaboration:2009bp}. In Figure \ref{fig:lowq2} they are
compared to phenomenological models based on the dipole approach
and good agreement is found.
At medium $Q^2$ up to $150\,\mathrm{GeV}$ the precision reaches
$1.3-2\%$ \cite{Collaboration:2009kv}. A NLO QCD fit (H1PDF2009) of
the complete H1 data collected up to the year 2000 is performed and
parton densities are extracted. The new data are described very well by the
QCD fit. The extracted parton densities 
have much reduced errors, as compared to earlier fits.
The new data on the structure function $F_2$ and the results of
the H1PDF2009 fit are shown in Figure \ref{fig:inclusive}.
Three error
sources are indicated on the fit results: the innermost error band are
experimental errors, followed by theory uncertainties. The outermost
error band includes the uncertainty due the PDF parametrisation,
which is defined here for the first time.

\begin{figure}[t]
\begin{center}
\includegraphics[width=0.49\columnwidth]{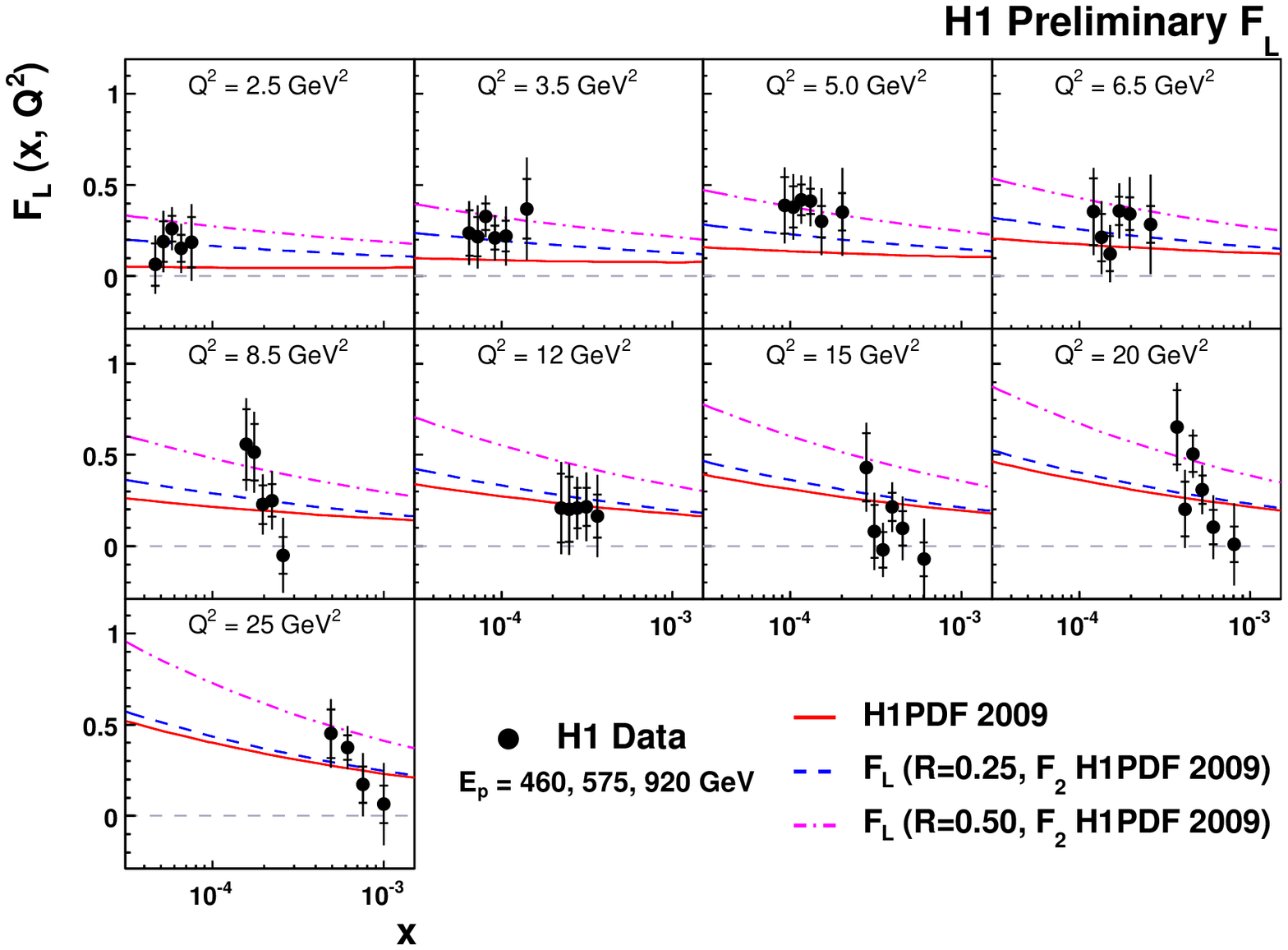}
\includegraphics[width=0.49\columnwidth]{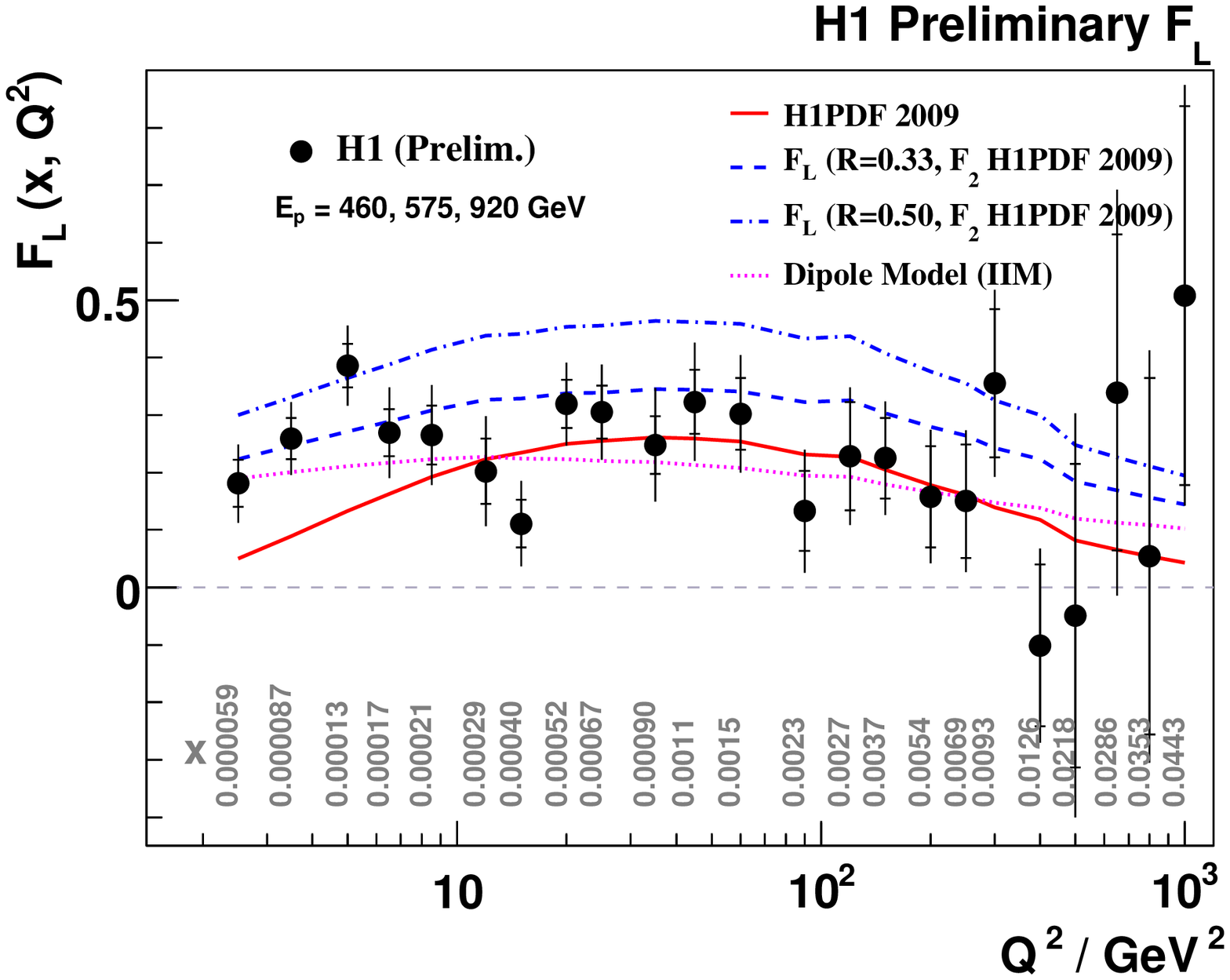}
\caption{The longitudinal structure function $F_L$,
  as a function of $x$ and $Q^2$ (left) and as a function of $Q^2$,
  averaged over the corresponding $x$ values (right)}\label{fig:fl}
\end{center}
\end{figure}
Using the data collected in dedicated low energy runs, a direct
measurement of the longitudinal structure function $F_L$ is
performed. Data collected at medium $Q^2$ \cite{:2008tx} are combined
with measurements at lower and higher $Q^2$ and together cover the kinematic
range $2.5\,Q^2<800\,\text{GeV}^2$. 
The results are shown in Figure \ref{fig:fl}. The new data at low
$Q^2$ are shown as a function of $x$ and $Q^2$. All H1 data for a given
$Q^2$ are then averaged over $x$ and the result is shown as a function
of $Q^2$. The data are compared to predictions of the H1PDF2009 QCD
fit and to a dipole model. At small $Q^2$ the $F_L$ predicted from the
QCD fit is lower than observed, whereas the dipole model is found to
be in agreement with the data. The data has a good potential to
constrain the models of the proton structure at low $x$.

\section{Heavy flavour production}

Charm and beauty quarks are produced at HERA predominantly in the
boson gluon fusion process, where the virtual photon and a gluon from
the proton couple to a heavy quark-antiquark pair. Heavy meson
production thus depends on the gluon content of the proton.
As the heavy quark mass is non-negligible compared to the 
scale $Q$, set by the exchanged virtual photon, theoretical prediction
need to include the quark masses. Measurements of the contributions
from heavy quarks to the inclusive structure function, $F_2^c$ and
$F_2^b$ hence provide both interesting tests of QCD and sensitivity to the
gluon density.
\begin{figure}[b]
\begin{center}
\includegraphics[height=0.66\columnwidth]{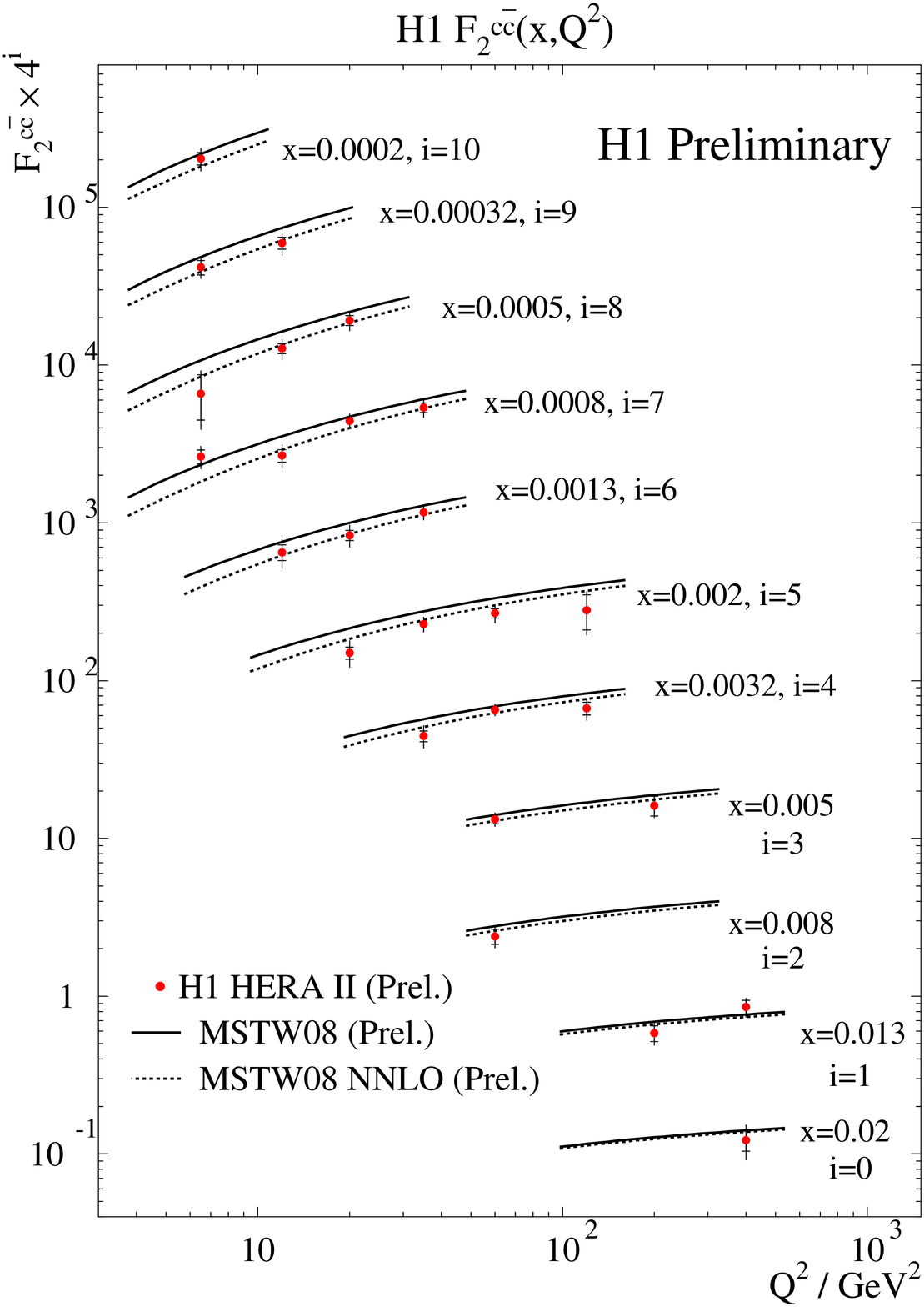}
\includegraphics[height=0.66\columnwidth]{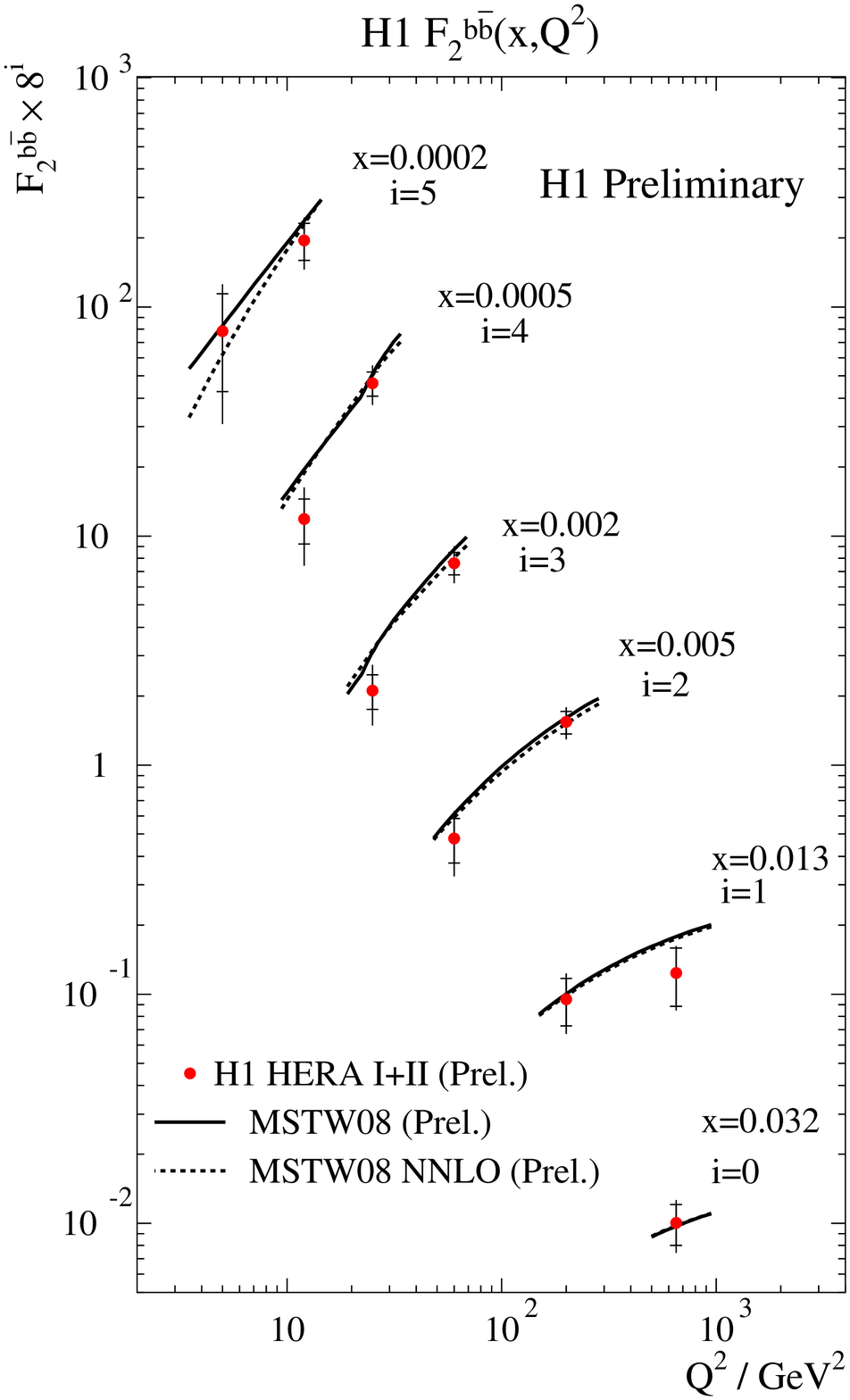}
\caption{Contribution from charm and beauty production to the
  structure function $F_2$, measured with a lifetime analysis}
\label{fig:f2ccbb}
\end{center}
\end{figure}
Figure \ref{fig:f2ccbb} shows these contributions measured using a
lifetime analysis: in addition to the scattered electron, tracks or
vertexes significantly displaced from the primary vertex are
required. The analysis covers the kinematic range
$5<Q^2<650\,\text{GeV}$. NLO QCD predictions are in
agreement with these measurements.

\begin{figure}[t]
\begin{center}
\includegraphics[height=0.48\columnwidth]{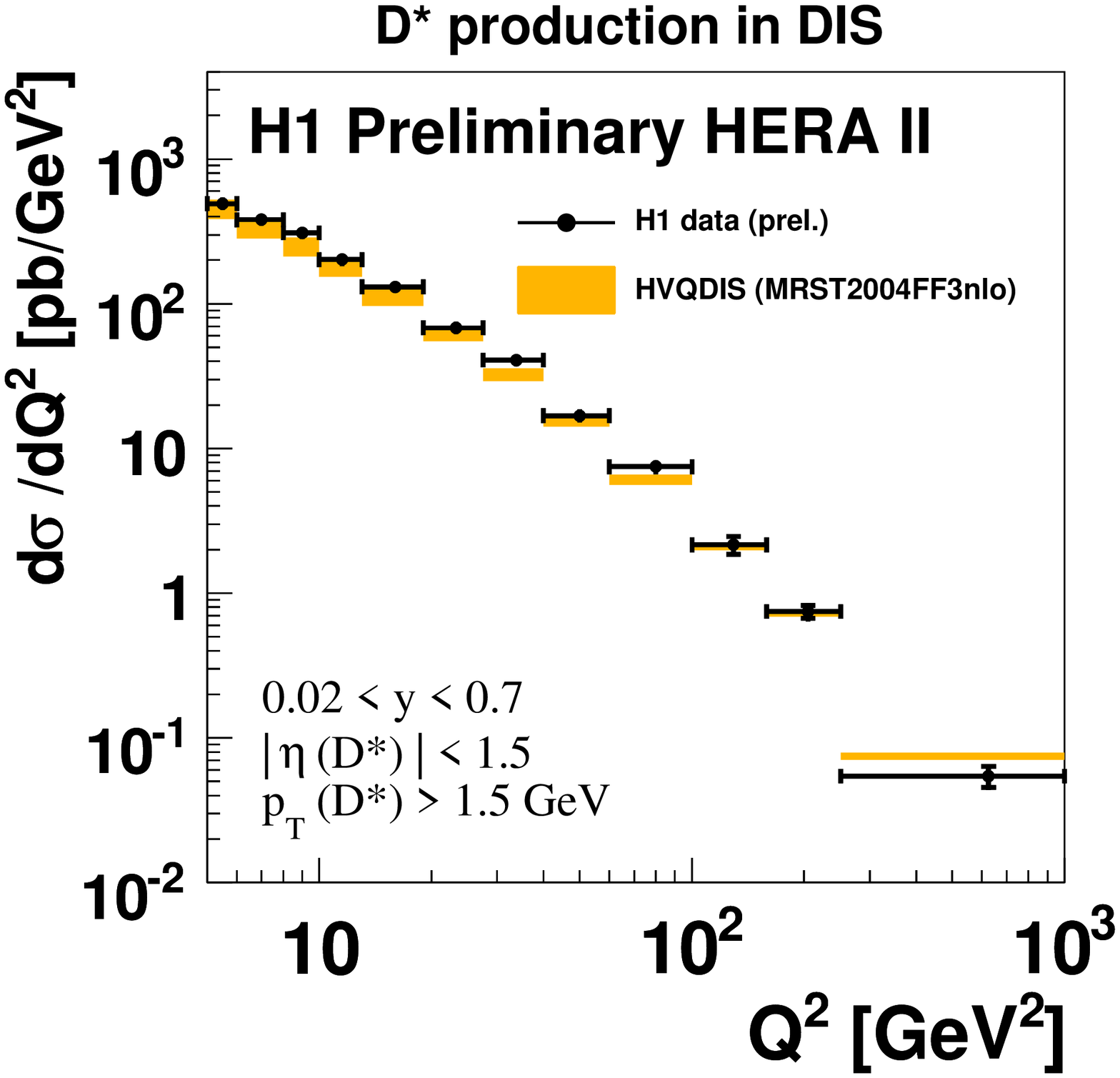}
\includegraphics[height=0.48\columnwidth]{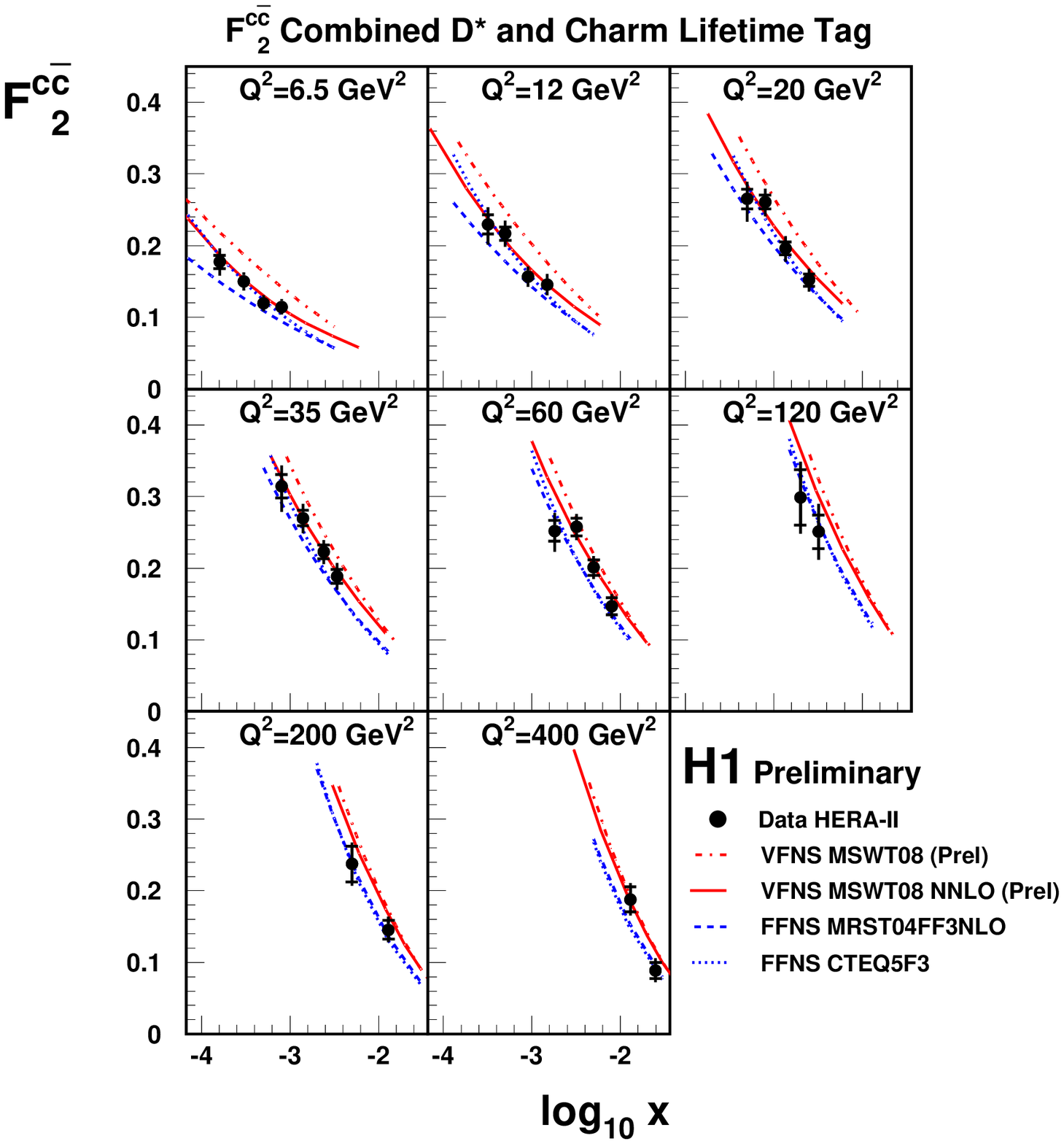}
\caption{Cross-section for $D^{\star}$ meson production (left) and the
  structure function $F_2^c$, after averaging $D^{\star}$ and lifetime
  data (right)}
\label{fig:dstar}
\end{center}
\end{figure}

Another way to experimentally access the charm contribution is by
reconstructing $D^{\star}$ mesons. The production of $D^{\star}$
mesons is measured in the kinematic range
$5<Q^2<800\,\text{GeV}$. Good agreement with calculations is found,
and the measurement is extrapolated to the full phase-space. Figure
\ref{fig:dstar} shows the cross-section for $D^{\star}$ as a function
of $Q^2$ and the structure function $F_2^c$ after combining 
the $D^{\star}$ data with the lifetime analysis.
The structure function $F_2^c$ is compared to QCD predictions based on
different parameterisations of the parton densities. The QCD predictions are
found to be in agreement with the combined $F_2^c$ data. At low $Q^2$ the data
start to discriminate between various schemes of including heavy quark masses
in the theory. The precision of the data is comparable with the spread
in the theoretical predictions and prefigures further constraints
on the proton structure.

\section{Measurements of the hadronic final state}

The production of jets at HERA provides a clean environment to study 
fundamental properties of QCD, like the strong coupling constant
$\alpha_s$. The H1 collaboration measures inclusive jet production
normalised to inclusive DIS, as well as the rate of 2-jet and 3-jet
events \cite{Collaboration:2009vs}. The measurements are performed as
a function of $Q^2$ and $\langle P_T\rangle$ in the kinematic range
$150<Q^2<15000\,\text{GeV}^2$, where $\langle P_T\rangle$ is the
jet transverse momentum. A value
$\alpha_s=0.1168\pm0.0007\text{(exp.)}^{+0.0046}_{-0.0030}\text{(theor.)}\pm0.0016\text{(PDF)}$
is found in an NLO QCD analysis. The small experimental error on $\alpha_s$ compared to the dominant
theoretical error emphasises the need of higher order calculations for
jet production at HERA.
\begin{figure}[t]
\begin{center}
\includegraphics[height=0.4\columnwidth]{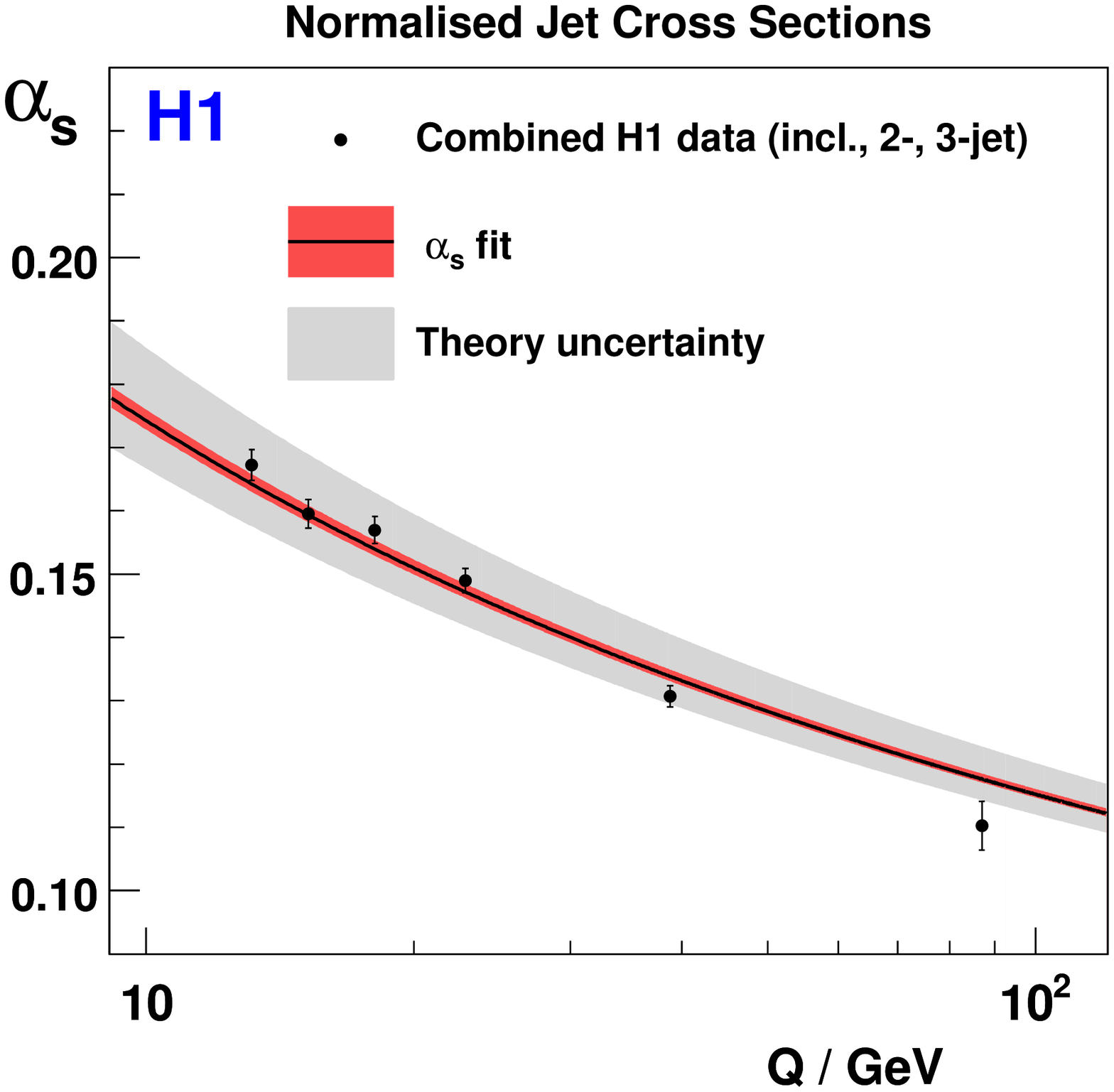}
\includegraphics[height=0.4\columnwidth]{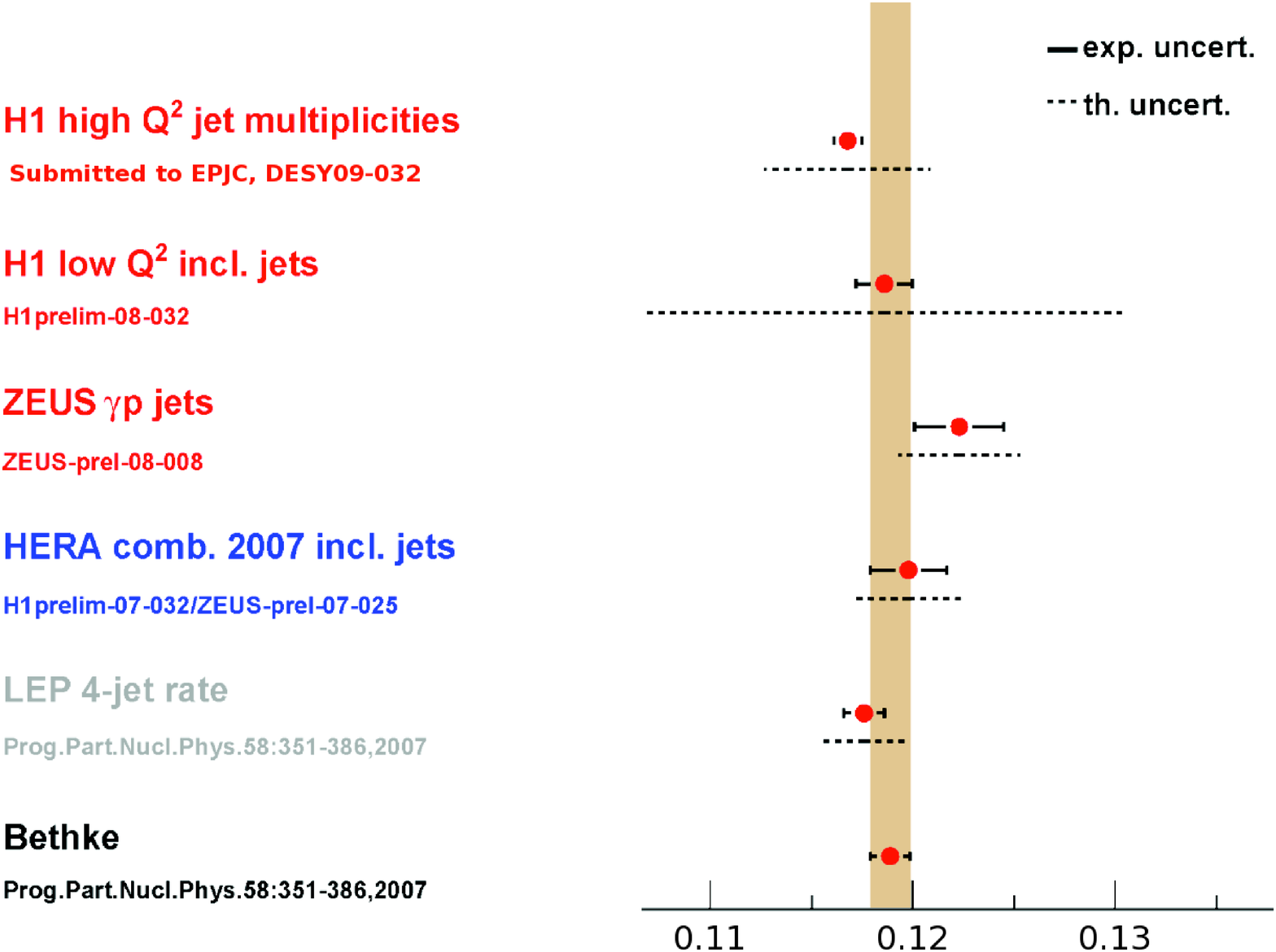}
\caption{Measurements of $\alpha_s$ from jets at HERA}
\label{fig:alphas}
\end{center}
\end{figure}
\begin{figure}[t]
\begin{center}
\includegraphics[width=0.605\columnwidth]{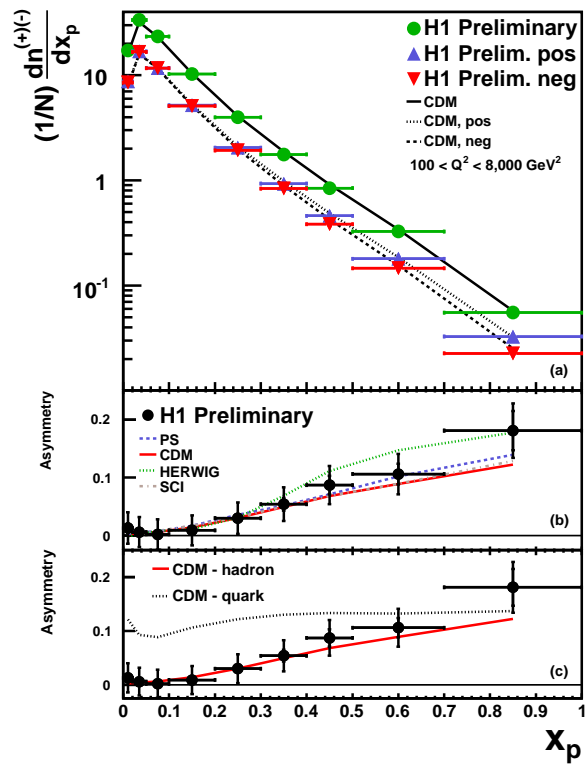}
\caption{The particle charge asymmetry measured at HERA}\label{fig:cpa}
\end{center}
\end{figure}
The running of $\alpha_s$ with $Q^2$, as seen in jet production at HERA and a
comparison to other $\alpha_s$ measurements is shown in Figure
\ref{fig:alphas}.

As a further investigation of QCD effects in the hadronic final state
structure, the production of charged particles is studied.
The charged particle spectra are sensitive to the fragmentation
process, in which hadrons are produced from the initial state
partons. As opposed to the hard process, which can be calculated in
perturbation theory, fragmentation is described by phenomenological models.
H1 measured the production of charged particles \cite{Aaron:2007ds} as a
function of $x_p$, the momentum fraction in the current hemisphere of the
Breit frame, in the kinematic range $100<Q^2<20000\,\text{GeV}$. More insights
in the fragmentation process are obtained by studying particle production
separately for negatively and positively charged particles. These rates and
the associated asymmetry are shown in Figure \ref{fig:cpa}.
The asymmetry is small
at low $x_p$ and rises up to $0.18$ for high $x_p$, in agreement with various
fragmentation models. In contrast, the charge asymmetry observed at parton
level, does not depend strongly on $x_p$.

\section{Diffraction}

\begin{figure}[b]
\begin{center}
\includegraphics[height=0.7\columnwidth]{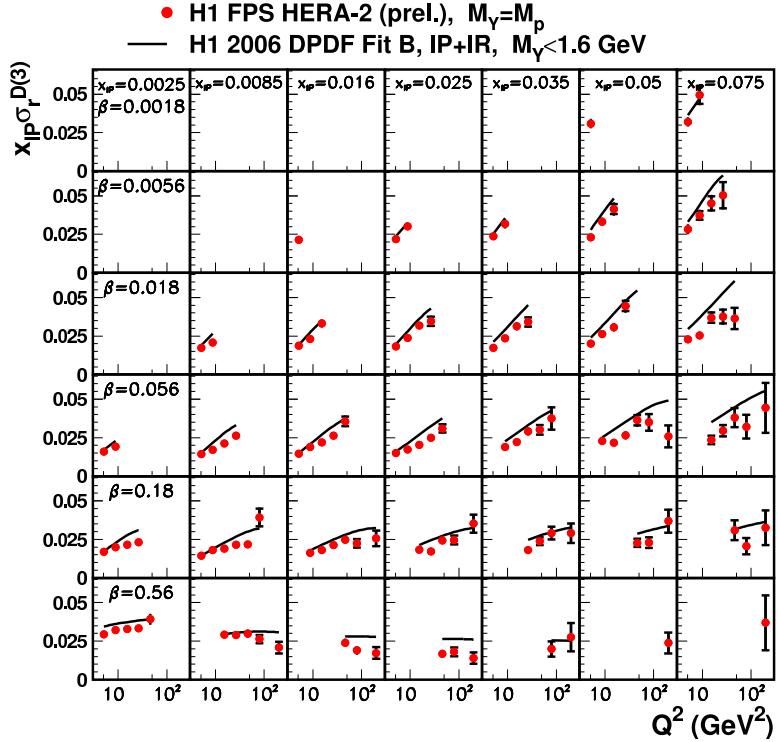}
\caption{Leading proton data on $x_{\text{I\-P}}\sigma_R$}
\label{fig:fps}
\end{center}
\end{figure}
Diffractive processes contribute at a level of about $10\%$ to the DIS
cross-sections at HERA. They are characterised by the fact
that the proton stays intact or fragments into a system with small mass
$M_Y$, even though a second hadronic system is produced. This is interpreted
as the exchange of a colour-neutral hadronic object. In leading order QCD,
this is explained by the exchange of two gluons. Experimentally, diffractive
events are detected by identifying the outgoing proton or by requiring a large
rapidity gap free of hadronic activity and separating the two hadronic
systems.  New results on inclusive diffraction are derived from data
taken with the H1 Forward Proton Spectrometer (FPS). Leading protons
are detected in the FPS, which is located about $80\,\text{m}$
downstream the main H1 detector in the HERA tunnel, using roman pot
detectors inside the evacuated part of the beam transport
system. Compared to earlier measurements with a leading proton
\cite{Aktas:2006hx}, the analysed data sample corresponds to an
integrated luminosity increased by one order of magnitude.
The resulting reduced cross-section is shown in Figure
\ref{fig:fps} as a function of three variables: $x_{\text{I\-P}}$, $\beta$ and
$Q^2$. The variable $\beta$ is related to Bjorken $x$ and the momentum
fraction $x_{\text{I\-P}}$ of the colourless object entering the hard
interaction, $\beta=x/x_{\text{I\-P}}$. The data are compared to
predictions from a QCD fit to inclusive H1 rapidity gap data
\cite{Aktas:2006hy}. Agreement is found, taking
into account the fact that the fit includes data up to a mass
$M_Y<1.6\,\mathrm{GeV}$, whereas the new data are collected with a
leading proton, i.e $M_Y=M_p$. Compared to \cite{Aktas:2006hx}, the
new measurements extend to higher $Q^2$. The precision reached is of
order $8\%$ at low $Q^2$, limited by systematic effects.

\begin{figure}[b]
\begin{center}
\includegraphics[height=0.35\columnwidth]{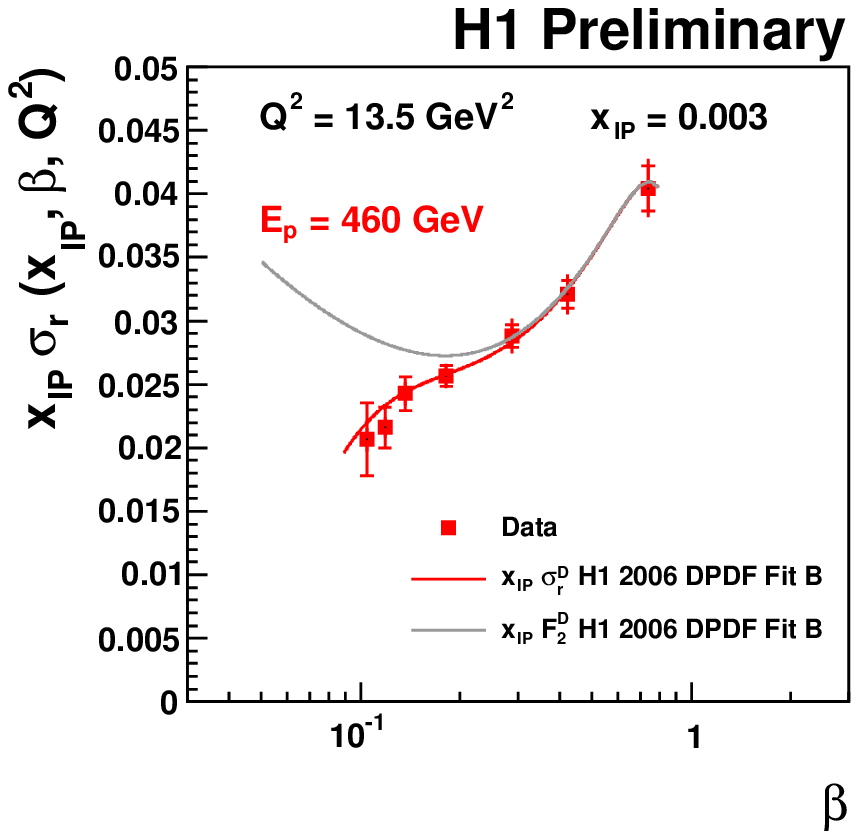}
\includegraphics[height=0.35\columnwidth]{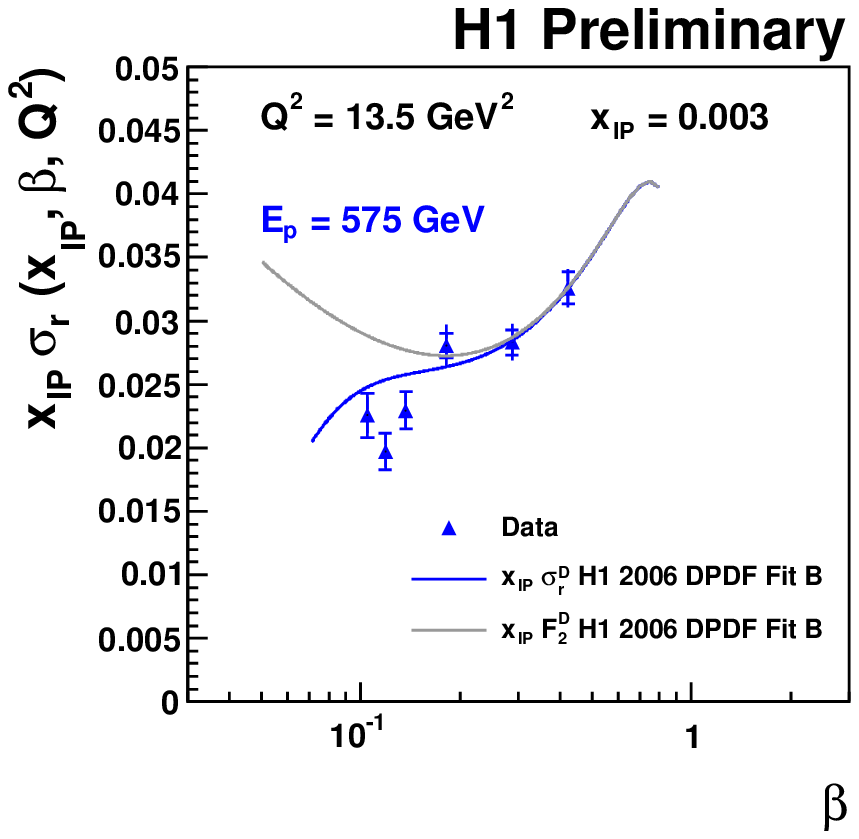}
\includegraphics[height=0.35\columnwidth]{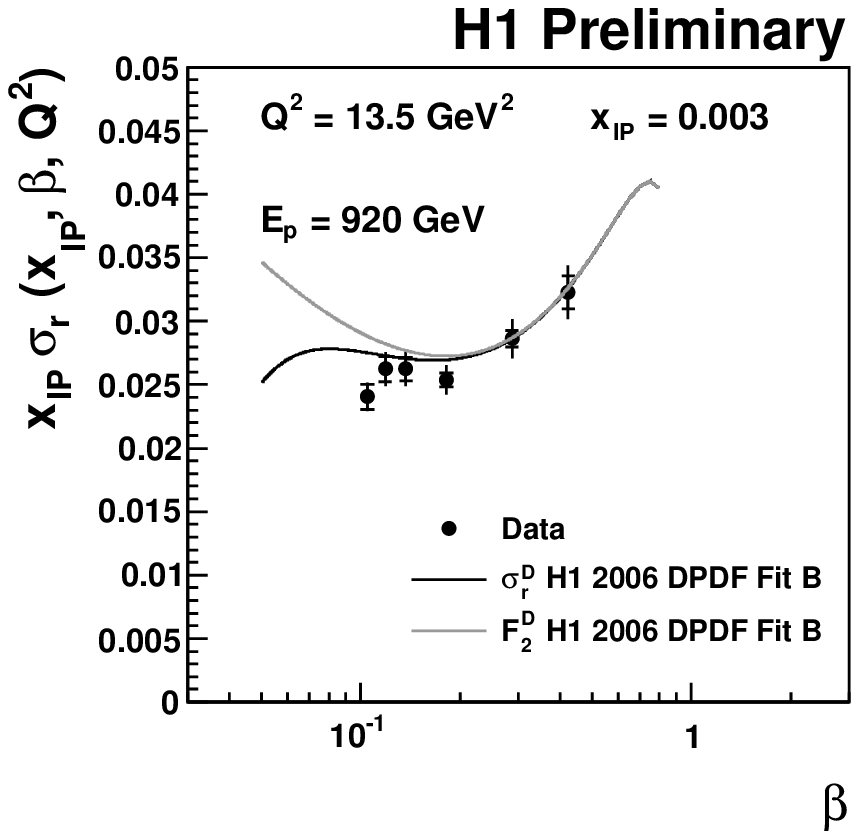}
\includegraphics[height=0.35\columnwidth]{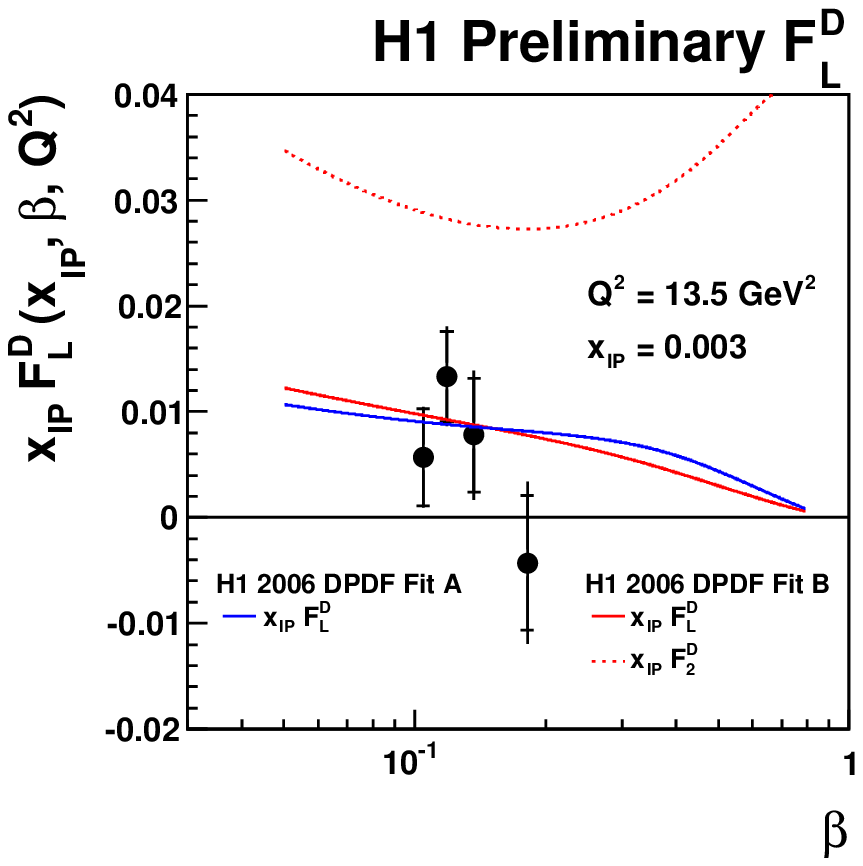}
\caption{The diffractive reduced cross-section $x_{\text{I\-P}}\sigma_r$
  measured for three centre-of-mass energies and the extracted 
  structure function $F_L^D$ (lower left)} \label{fig:fld}
\end{center}
\end{figure}
A first measurement of the diffractive longitudinal structure function
$F_L^D$ is performed, using data collected in the low energy runs. 
The diffractive reduced cross section
$x_{\text{I\-P}}\sigma_r$ is decomposed into a linear combination of the
structure functions $F_2^D$ and $F_L^D$. The contribution from $F_2^D$ is
independent of the centre-of-mass energy, in contrast to the
contribution proportional to $F_L^D$. Thus, by analysing
$x_{\text{I\-P}}\sigma_r$ at various beam energies, $F_L^D$ is extracted. The
quantity $x_{\text{I\-P}}\sigma_r$ for various beam energies and the
resulting structure function $F_L^D$ are shown as a function of
$\beta$ in Figure \ref{fig:fld}. 
The $F_L^D$ data agree with predictions derived from the diffractive PDF fit
\cite{Aktas:2006hy} to inclusive H1 data.

\begin{figure}[b]
\begin{center}
\includegraphics[height=0.49\columnwidth]{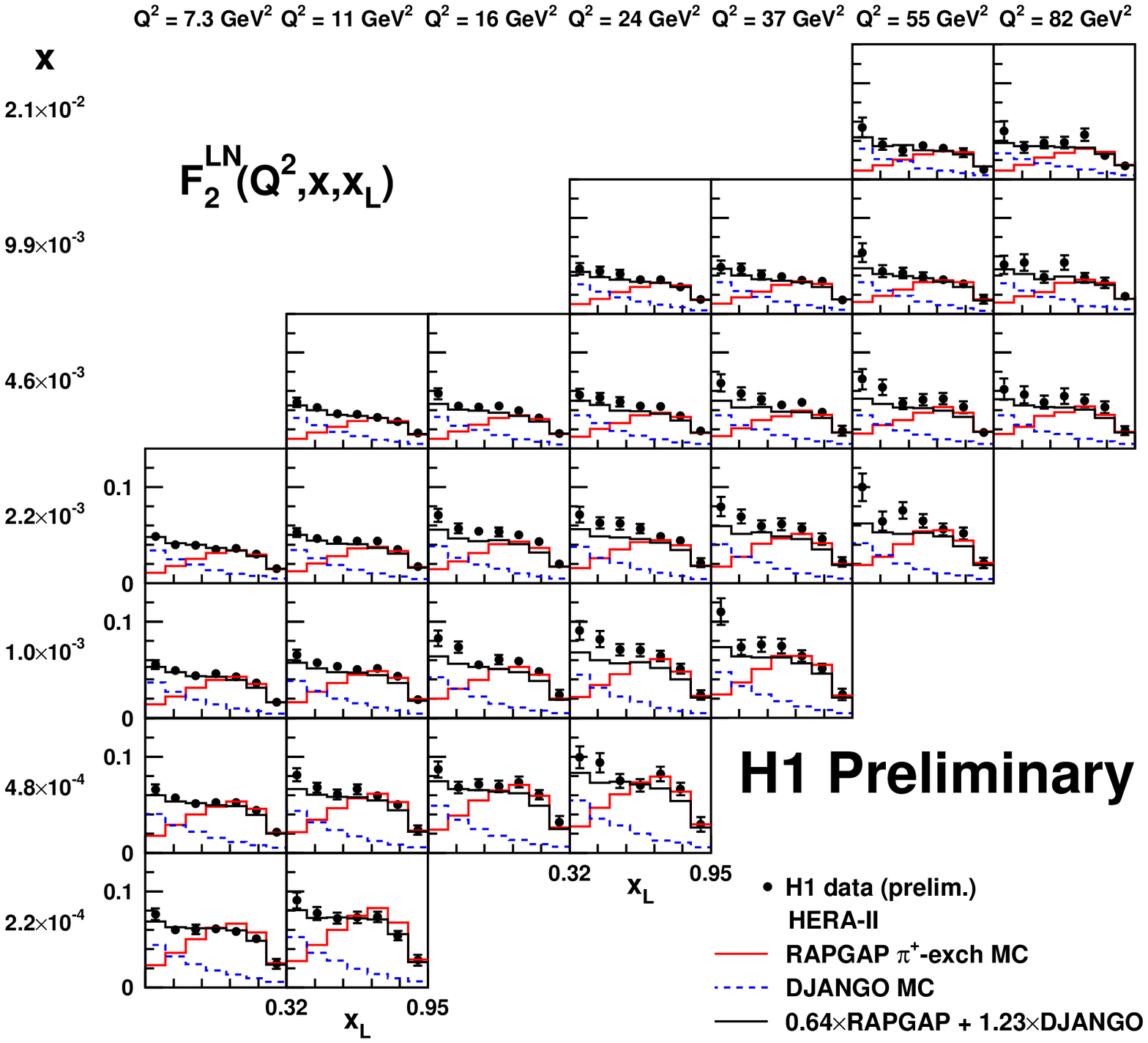}
\includegraphics[height=0.49\columnwidth]{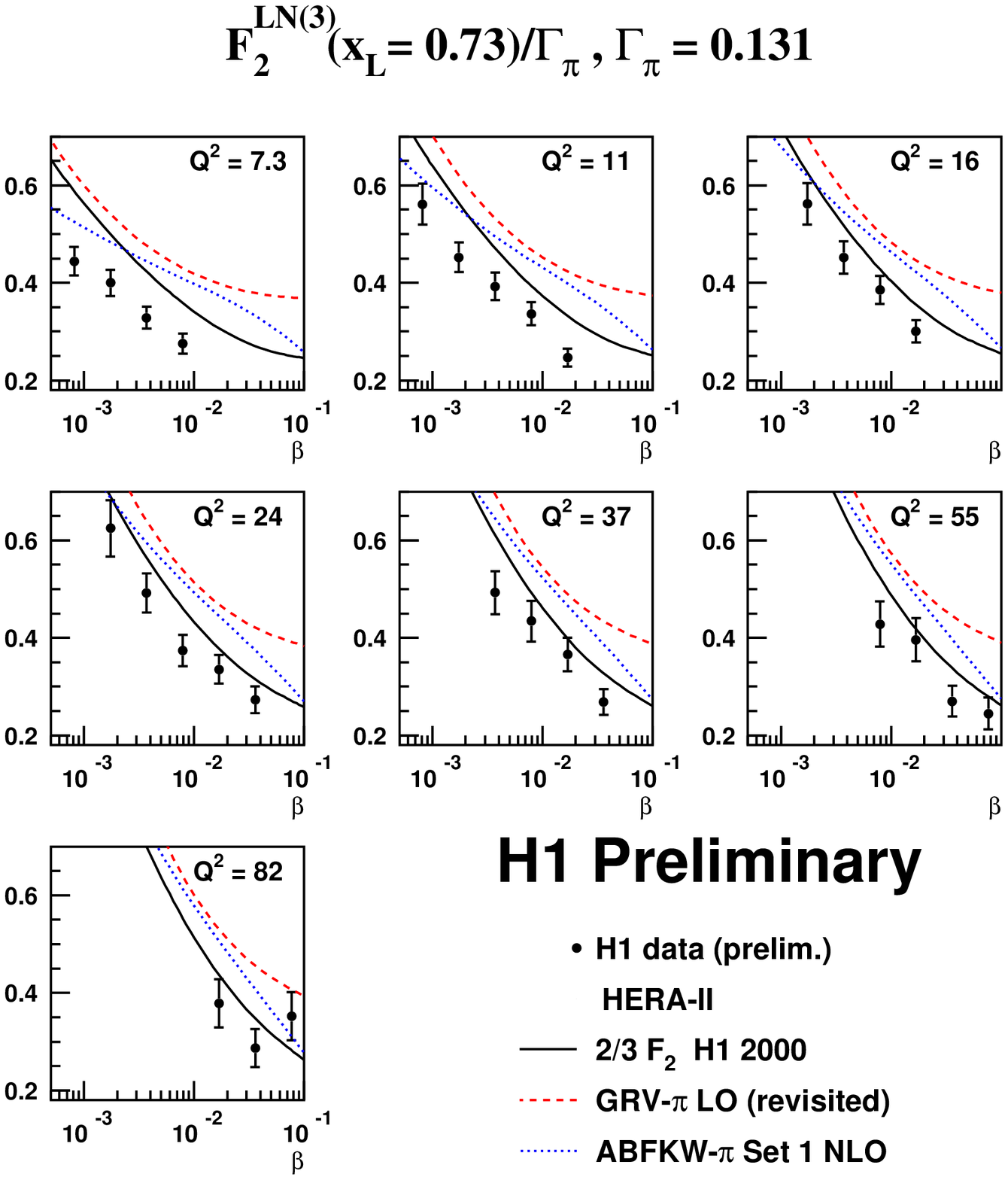}
\caption{The structure function $F_2^{LN}$ with a leading neutron
  (left) and comparisons to parameterisations of the pion structure
  function (right)}
\label{fig:f2ln}
\end{center}
\end{figure}

It is also interesting to look for the production of leading neutrons,
which are produced from an electrically charged, colourless exchange
with the proton. Such neutrons are detected in the Forward Neutron
Calorimeter (FNC) located $106\,\text{m}$ from the main detector in
the HERA tunnel. Compared to earlier measurements
\cite{Adloff:1998yg}, the new data profit from much increased integrated 
luminosity and an improved FNC. Figure \ref{fig:f2ln} shows the
measured structure function $F_2^{LN}$ as a function of the three variables
Bjorken $x$, $Q^2$ and $x_L$, where $x_L$ is the neutron energy, normalised to
the proton beam energy.
The data are compared to predictions from two Monte Carlo (MC) models: DJANGO, where
leading neutrons are produced in the fragmentation process, and RAPGAP
$\pi^{+}$ where leading neutrons are produced by the exchange of a
charged pion. After applying global normalisation constants, the sum of these MC
models is able to describe the data.
Also shown in Figure \ref{fig:f2ln} are the $F_2^{LN}$ data at $x_L=0.73$
divided by a corresponding pion flux factor $\Gamma_{\pi}=0.131$.
At high $x_L$ the data are dominated by the pion exchange, and hence the pion
structure function may be extracted. The $F_2^{LN}$ data divided by
the pion flux factor compare well in shape to various parameterisations of the
pion structure function.

\section{Summary}

The H1 collaboration has presented a wealth of new results based on the
datasets collected until mid 2007. A selection of
these results in the areas searches for new physics, structure functions,
heavy flavour production, hadronic final states and diffraction are 
discussed.

Most searches have now been completed, using the full HERA
statistics. The combination of H1 and ZEUS data has started, in order to reach
the best possible sensitivity to physics beyond the standard model.

The precision of inclusive cross-sections data measured by H1 is now
as good as $1.3-2\%$ in the bulk of the 
phase-space, opening new possibilities for QCD fits. The structure function $F_L$
is measured over the full $Q^2$ range available at HERA. Of particular
interest is the region of lowest $Q^2$, where the models diverge most and the
data start to discriminate amongst them.

Using the H1 vertex-detector, contributions from charm and beauty to the
inclusive structure function $F_2$ are extracted. In combination with
data from $D^{\star}$ decays, the data show sensitivity to the heavy flavour
treatment in QCD fits and may contribute to further constrain the
proton structure.

New measurements of jet production at high $Q^2$ lead
to very small experimental errors on the strong coupling $\alpha_s$. Higher
order calculations are needed to pin down the theory errors. The
charged particle asymmetry of the scattered hadronic final state is
observed for the first time in $ep$ collisions and found in agreement
with predictions based on phenomenological models.

A measurement of the diffractive longitudinal structure function $F_L^D$
is presented here for the first time. Diffractive PDF fits to
inclusive H1 data are in agreement with this new data. Using dedicated
detectors FPS and FNC the production of leading protons and leading
neutrons, respectively, is measured. The new leading proton data
extend the kinematic range and have superior precision as compared to
earlier measurements. They are in agreement with predictions of the
diffractive PDF fit. The leading neutron data can be used to test
fragmentation models and to extract the pion structure function.

The results presented here reveal the potential of the full HERA
physics data, which will continue to be exploited in the next years.



\begin{footnotesize}

\end{footnotesize}


\end{document}